\documentstyle[12pt]{article}
\begin{document}
\begin{titlepage}
\begin{flushright} UTPT-93-26 \\ hep-ph/9310216 \\ September 1993
\end{flushright}
\vspace{24pt}
\begin{center} {\LARGE Comparison of $1/{m}_{Q}^{2}$ Corrections\\in Mesons
and Baryons}\\ \vspace{40pt} {\large B.
Holdom\footnote{holdom@utcc.utoronto.ca}, M.
Sutherland\footnote{marks@medb.physics.utoronto.ca} and J. Mureika}
\vspace{0.5cm}

{\it Department of Physics\\ University of Toronto\\
Toronto, Ontario\\Canada M5S 1A7}

\vspace{12pt}

\begin{abstract} We extend our relativistic quark model to the study of the
decay
${\Lambda }_{b}\rightarrow{\Lambda }_{c}\ell\overline{\nu}$ and verify that
the model satisfies the heavy-quark symmetry constraints at order
$1/{m}_{Q}^{2}$.  We isolate a strong dependence on a parameter which
measures the relative distortion in the light-quark wave functions of the
${\Lambda }_{b}$ and ${\Lambda }_{c}$.  This parameter and the
$1/{m}_{Q}^{2}$ corrections turn out to be small. The dependence on a
corresponding parameter in the meson case leads to large $1/{m}_{Q}^{2}$
corrections.
\end{abstract}
\end{center}  \end{titlepage} \pagebreak
\setcounter{footnote}{0}

In previous work we have discussed by way of an explicit model the possible
effects of hyperfine splitting physics on semileptonic decays of heavy mesons
\cite{simply,hyperfine,linear}.  The model was expanded to first order in the
heavy-quark mass expansion and was found to be fully consistent with the
heavy-quark effective theory (HQET).  But the model could also be treated in
unexpanded form.  In this case we saw that corrections occurring at order
$1/m_Q^2$ and higher could shift the implied value of $V_{cb}$ by 10 or 15\%.

If predictions from first-order HQET are to be believed, it is important to
be able to confirm or rule out these possible large corrections at higher
orders.  One would hope that if these corrections were actually present they
would give rise to other experimental signatures.  But our explicit meson
model provides an example in which other effects of the hyperfine
corrections are quite well hidden.  In fact we find that in the semileptonic
decays
$B\rightarrow {D}^{*}$ and $B\rightarrow D$ it is quite possible that
there is no other observable effect besides the shift in $V_{cb}$.  In
particular we find that all other predictions inherent in first-order HQET
are satisfied to good approximation.  Some examples of this result may be
found in \cite{linear}.

There are six form factors, four (${h}_{V}$, ${h}_{{A}_{1}}$,
${h}_{{A}_{2}}$, ${h}_{{A}_{3}}$) for $B\rightarrow {D}^{*}$ and two
(${h}_{+}$,${h}_{-}$) for $B\rightarrow D$.  It is only the products of each
of these form factors with $V_{cb}$ which are observable in these decays,
i.e. a scaling of the form factors by a common factor and a scaling of
$V_{cb}$ by the inverse factor is not observable in these decays.  The form
factors of our model can be artificially rescaled in this way to make them
consistent with first-order HQET.  That is, after they are all scaled down
by 10 or 15\% they would then satisfy the normalization conditions at zero
recoil implied by first-order HQET. (But see the caveat below.)

In addition we find that these rescaled form factors can be reproduced to
good approximation by choosing a suitable Isgur-Wise function, along with a
suitable choice of the four universal functions appearing at first order in
HQET.\footnote{The determination of these functions as a ``best fit" to the
unexpanded model leads to results which are actually quite similar to those
found in the expanded version of the model.}  All experimental observables
would then be consistent with first-order HQET.  Thus in a ``world" described
by our model, it would be very tempting to extract $V_{cb}$ from the ``data"
assuming first-order HQET.  But this procedure would yield a result which was
incorrect by the amount of the scaling factor.  The use of the true unscaled
form factors from the model would yield a value for $V_{cb}$ a factor of 10
or 15\% smaller.

The caveat is that the rescaled form factors are compatible with first-order
HQET only if the model produces similar values for two of the form factors
at zero recoil, ${h}_{+}(1)$ and ${h}_{{A}_{1}}(1)$.  It is only in this
case that a common rescaling can approximately reproduce the first-order
HQET result ${h}_{{A}_{1}}(1)={h}_{+}(1)=1$.  The question of whether
${h}_{+}(1)$ and ${h}_{{A}_{1}}(1)$ are similar in the model is a question
of what $b$ and $c$ quark masses are used \cite{linear}.  It is conceivable
that these quark masses are such that ${h}_{+}(1)$ and ${h}_{{A}_{1}}(1)$
are quite different; this then produces an experimental signature.  In this
case, the $V_{cb}$ determined from $B\rightarrow {D}^{*}$ would not match
with the $V_{cb}$ determined from $B\rightarrow D$ decays, if first-order
HQET was assumed in both cases.  This could then be taken as evidence that
large corrections beyond first-order HQET were present.

The origin of the corrections lies in the fairly large hyperfine distortions
of the light-quark ``wave functions" for the $D$ and ${D}^{*}$ mesons due to
the finite mass of the charm quark.\footnote{Light-quark refers to all light
degrees of freedom.}  It is the relative distortion between these wave
functions and the less-distorted $B$-meson light-quark wave function which
produces the large corrections.  In the language of HQET we found
\cite{hyperfine} a large $1/m_c m_b$ correction originating from one
insertion of a chromomagnetic moment operator on each of the charm and bottom
quark lines.

In the model the amount of the relative wave function distortion can be
altered by holding the meson masses fixed and varying the $b$ and $c$ quark
masses.  For the quark masses outside of some range, the resulting
corrections to semileptonic $B$ decay can become absurdly large.  In fact we
find that the model, for fixed meson masses, restricts the quark masses to
lie within some narrow wedge-shaped region in the ${m}_{b}$-${m}_{c}$ plane
\cite{linear}.  This constraint relation between the $m_c$ and $m_b$ masses
is of interest since it may be used to help pin down these masses and other
quantities appearing in heavy-quark effective theory.  And we stress that this
constraint is another reflection of the large hyperfine corrections; if
hyperfine effects are artificially turned off then the allowed region is
much larger and the constraint disappears.

It has been suggested \cite{LukSav} that constraints on masses and parameters
in HQET can be obtained from inclusive $D$ and $B$ semileptonic decays.  With
the input of the inclusive $D$ decay rate and the physical $B$ and $D$ masses
there are three relations among the four quantities $m_c$, $m_b$,
$\overline{\Lambda }$, and the parameter ${\lambda }_{1}$ appearing in the
heavy-quark expansion of the meson masses.  The additional relation between
$m_b$ and $m_c$ provided by our model is then enough to determine these four
quantities.  One such set consistent with the various experimental and
theoretical uncertainties discussed in \cite{LukSav} is ${m}_{b}=4.85$ GeV,
${m}_{c}=1.5$ GeV,
$\overline{\Lambda }=.5$ GeV, and ${\lambda }_{1}=.35$ GeV${}^2$.  In
\cite{LukSav} the input of the $B$ inclusive rate also constrains $V_{cb}$;
an independent determination of $V_{cb}$ from this method would, in
principle, provide another signature of large hyperfine effects.  This is
because
$V_{cb}$ from the inclusive rate would differ from that obtained from the
exclusive decays, if only first-order HQET was used in the latter.

In this paper we will extend our model to cover the case of heavy-baryon
${\Lambda }_{b}\rightarrow{\Lambda }_{c}\ell\nu$ decays.  This case, in
which hyperfine splitting effects are absent, will provide an instructive
contrast to semileptonic $B$ decays.  In particular we need to confirm that
the higher-order corrections, now in the absence of hyperfine effects, are
in fact greatly reduced.  ${\Lambda }_{b}$ decays would then provide yet
another way to check the value of $V_{cb}$ obtained from $B$ decays.

Because of simplifications in our model for the baryon case we will be able
to perform the expansion analytically to order $1/m_Q^2$, one order higher
than in the meson case.  We verify that the model satisfies the heavy-quark
symmetry relations at this order.  We are then able to isolate a strong
dependence on a parameter which measures the amount of distortion in the
light quark wave functions.  This is analogous to the phenomenon
in the meson case which, we have argued, is responsible for the large
corrections there.

Denote by $\Lambda_{Q}$ the ground-state baryon with flavor content $Qud$
and light degrees of freedom having zero spin.  We model the light degrees
of freedom as a scalar such that the vertex between baryon,
heavy quark and scalar is
\begin{equation}
\frac{Z^2}{-k^{2}+\widehat{\Lambda }^{2}},
\end{equation} where $k$ is the momentum of the scalar.  This is a damping
factor which prevents large momentum from flowing into the light degrees of
freedom.  All our results follow from one-loop graphs in which we use standard
propagators for the scalar with mass $m$ and the heavy quark with mass
$m_{Q}$.

Let $\Sigma(p\!\!\!/)$ be the baryon self-energy. The relation Re
$\Sigma(M)=0$, where the physical baryon mass $M$ is real, fixes
$\widehat{\Lambda }$ in terms of $M$, $m_{Q}$ and $m$.  The normalization
factor
$Z$ is fixed in terms of $M$, $m_{Q}$ and $m$ by the condition Re
$\Sigma^{\prime}(M)=1$, where the prime denotes differentiation with respect
to $p\!\!\!/$.  Thus the only parameters of the unexpanded model are the
heavy-quark and scalar masses.

Form factors for $\Lambda_{Q} \rightarrow \Lambda_{Q^{\prime}}$ are defined
by \cite{IsgWis} \begin{eqnarray} \langle \Lambda_{Q^{\prime}} |
	\overline{Q}^{\prime}\gamma_{\mu}Q | \Lambda_{Q} \rangle  &=&
\overline{u}^{\prime} [ F_{1}\gamma_{\mu} + F_{2}v_{\mu} +
F_{3}v^{\prime}_{\mu}] u \nonumber \\ \langle \Lambda_{Q^{\prime}} |
	\overline{Q}^{\prime}\gamma_{\mu}\gamma_{5}Q | \Lambda_{Q} \rangle  &=&
\overline{u}^{\prime} [ G_{1}\gamma_{\mu} + G_{2}v_{\mu} +
G_{3}v^{\prime}_{\mu}]\gamma_{5} u \end{eqnarray} where $v$ and $v^{\prime}$
are the initial and final four-velocities and the spinors are normalized as
$\overline{u}u=2M$.  We take the form factors to be functions of
$\omega=v\cdot v^{\prime}$.  Vector current conservation implies that
$F(1)\equiv F_{1}(1) +F_{2}(1)+F_{3}(1)=1$ when $Q=Q^{\prime}$.

In the expanded version of our model we expand all quantities up to ${\cal
O}(1/m_{Q}^2)$ and ignore higher-order terms. \begin{eqnarray} \widehat{\Lambda
}
&=&\Lambda\{1 + \lambda ({\Lambda}/{m_Q}) +
\lambda^{\prime}{({\Lambda}/{m_Q})}^2\} \label{a0} \\ (M^2 -
m_{Q}^2)/\Lambda m_Q &=&c\{1 + d ({\Lambda}/{m_Q})
+d^{\prime}{(\Lambda/m_Q)}^2 \} \label{a1} \\
 Z/\Lambda&=&A\{1+B(\Lambda/m_Q)+B^{\prime} ({\Lambda}/{m_Q})^2 \}.
\end{eqnarray}  The quantities $\widehat{\Lambda }$ in Eq. (\ref{a0}) may be
different for different $Q$ but have a common value $\Lambda$ in the heavy
quark limit.  The parameters $\lambda$ and $\lambda^{\prime}$, which
characterize the approach to this limit, will determine the relative wave
function distortion between the $\Lambda_b$ and $\Lambda_c$.  The
zeroth-order quantities $c$ and $A$ and are determined by the ratio
$\alpha=m/\Lambda$, the first-order quantities $d$ and $B$ are determined by
$\alpha$ and $\lambda$, and the second-order quantities $d^{\prime}$ and
$B^{\prime}$ are determined by $\alpha$, $\lambda$ and $\lambda^{\prime}$.

For the form factors the expansions are \cite{FalkNeu} \begin{eqnarray}
F_{1}&=&\zeta+(\varepsilon^{\prime}+\varepsilon) [\zeta/2+\chi]
+(\varepsilon^{\prime 2}+\varepsilon^2)[b_{1}-b_{2}]
+\varepsilon^{\prime}\varepsilon[b_{3}-b_{4}] \nonumber \\
F_{2}&=&-\varepsilon^{\prime}\zeta/(\omega+1)  + \varepsilon^{\prime 2} b_2
+ \varepsilon^{\prime}\varepsilon b_5 \nonumber \\
F_{3}&=&-\varepsilon\zeta/(\omega+1)  + \varepsilon^2 b_2 +
\varepsilon^{\prime}\varepsilon b_5 \nonumber \\
G_{1}&=&\zeta+(\varepsilon^{\prime}+\varepsilon)
[(\omega-1)\zeta/2(\omega+1)+\chi]  +(\varepsilon^{\prime
2}+\varepsilon^2)b_{1} +\varepsilon^{\prime}\varepsilon b_{3}
\nonumber \\ G_{2}&=&-\varepsilon^{\prime}\zeta/(\omega+1) +
\varepsilon^{\prime 2} b_2 + \varepsilon^{\prime}\varepsilon b_6 \nonumber \\
G_{3}&=&\varepsilon\zeta/(\omega+1) -
\varepsilon^2 b_2 - \varepsilon^{\prime}\varepsilon b_6 \label{2ndorder}
\end{eqnarray} where
$\varepsilon_{Q}=\overline{\Lambda}/m_{Q}$.\footnote{Note that $b_i$ of
\cite{FalkNeu} are $4\overline{\Lambda}^2$ times $b_i$ here.} The
zeroth-order mass difference $\overline{\Lambda}=M-m_{Q}=c\Lambda/2$ need
not be equal to its counterpart in mesons.  Similarly, the Isgur-Wise
function $\zeta$ is unrelated to the analogous function $\xi$ in mesons.
Vector current conservation implies $\zeta(1)=1$, $\chi(1)=0$ and
$2b_1(1)+b_3(1)-b_4(1)+2b_5(1)=0$.  The quantities $F(1)$ and $G_{1}(1)$ are
especially interesting since they receive no first-order corrections
\cite{GGW}.

The expansions in (\ref{2ndorder}) depend on a number of new
universal functions in a manner consistent with heavy-quark symmetries
\cite{GGW,FalkNeu}.  The symmetry constraints reduce, for example, the
eighteen possible functions at second order down to six.  We have verified
analytically that the model satisfies these constraints.

The zeroth-order constant $c$ is determined by the condition
\begin{equation} 0=1/D_k^2 D_{\alpha} D,
\end{equation} where $D_k = -k^2 +1$, $D_{\alpha} = -k^2 + \alpha^2$ and $D
= -2 k \cdot v -c$.  An overall  $4 i A^4 \int d^4k/(2\pi)^4$ is understood
in any product of $1/D$'s, and the real part is implied. With the same
convention, the zeroth-order constant $A$ is determined by
\begin{equation} 1=1/D_k^2 D_{\alpha} D^2. \label{A}
\end{equation} The Isgur-Wise function is given by
\begin{equation}
\zeta=1/D_k^2 D_{\alpha} D D^{\prime} ,
\label{isgwis}
\end{equation} where $D^{\prime}= -2 k \cdot v^{\prime} -c$.  The Isgur-Wise
function is equal to 1 at zero recoil $v=v^{\prime}$ by virtue of (\ref{A}).
For the first-order constants we find
\begin{eqnarray} c \, d &=& c^2/2-1
	+ 1/D_{k} D_{\alpha} D^2 + 1/4D_{k}^2
	D_{\alpha} + 4 \lambda/D_{k}^3 D_{\alpha} D  \\ B&=& c/4
- (k^2 - c^2/2 + c\,d)/2D_{k}^2 D_{\alpha}
D^3+\lambda/D_{k}^3 D_{\alpha} D^2 \label{B}
\end{eqnarray} and for the sole first-order universal function we have
\begin{equation} c\chi/2 = (2B-c/2)\zeta  +
	(k^2-c^2/2+c\,d)/D_{k}^2 D_{\alpha} D^2 D^{\prime}-2\lambda/D_{k}^3
D_{\alpha}D D^{\prime},
\end{equation}  which vanishes at zero recoil by virtue of (\ref{B}).  There
are analogous expressions for the second-order quantities $d^{\prime}$,
$B^{\prime}$ and $b_{i}$ which we omit for brevity.

In order to illustrate the numerical values of the various quantities we
take the heavy-quark masses ${m}_{b}=4.85$ GeV, ${m}_{c}=1.5$ GeV, as
deduced above in connection with meson semileptonic decays.  For the scalar
mass we take $m=500$ MeV, since an appropriate effective light quark mass in
the case of mesons was found to be $m_{q}=250$ MeV.

We consider the unexpanded model first and by using the physical masses of
the baryons we find the values of
$\widehat{\Lambda }$ and $Z$ as shown in Table \ref{table1}. \begin{table}
\begin{center} \begin{tabular}{ccccc}
              &  $M$  & $m_{Q}$ & $\widehat{\Lambda }$ &  $Z$  \\  \hline
$\Lambda_{b}$ & 5.641 &  4.850  &   1.023       & 1.436 \\
$\Lambda_{c}$ & 2.285 &  1.500  &   1.008       & 1.549
\end{tabular} \end{center}
\caption{Physical baryon masses, input quark masses and resulting unexpanded
values of
$\widehat{\Lambda }$ and $Z$ (in GeV), for $m=500$ MeV.\label{table1}}
\end{table} With these values, we find the form factors shown in Fig. 1. The
net deviations of $F$ and $G_{1}$ from $\xi$ are very small across the
spectrum. At zero recoil, where the first-order corrections vanish
identically, we find
$F(1)=.9970$ and
$G_1(1)=.9964$ in the unexpanded model.  The net corrections to the heavy
quark limit in these quantities are very small and negative, quite unlike
the corrections we found in analogous quantities in the meson case.

The source of this result lies in the fact that the values of
$\widehat{\Lambda }$ for the $\Lambda_{b}$ and $\Lambda_{c}$ are very
similar. This result is not sensitive to the scalar mass.  For example,
when $m=660$ MeV we find $\widehat{\Lambda }=907$ MeV and 894 MeV for
$\Lambda_{b}$ and $\Lambda_{c}$, respectively.  In the case of $B$ meson
decay, the $\widehat{\Lambda}$'s for the
$D$ and $D^*$ mesons could not simultaneously be made very close to the
$\widehat{\Lambda}$ for the $B$ meson because of the constraint imposed by
the hyperfine mass splitting between the $D$ and $D^*$.

In the expanded model we find that it is consistent for second-order
corrections to be small, in contrast to the case in mesons. From the fact
that the values of
$\widehat{\Lambda }$ are so similar for the $\Lambda_{b}$ and $\Lambda_{c}$ it
is natural for $\lambda$ and $\lambda^{\prime}$ to be small (see Eq.
(\ref{a0})), with $\Lambda$ close to 1 GeV.  For illustrative purposes, we
determine these parameters as follows:  for a given $\Lambda$, we solve for
$\lambda$ and $\lambda^{\prime}$ by setting the second-order expressions for
the squared baryon masses equal to their experimental values.  $\Lambda$ is
then varied so that $F(1)$ and $G_{1}(1)$ at second order are as close as
possible to their unexpanded values.  We find
$\Lambda=1.019$ GeV, which corresponds to $c=1.553$ and
$\overline{\Lambda}=791$ MeV.  This compares with $\overline{\Lambda}=500$
MeV in the case of mesons. The optimal values of the other quantities are
$\lambda=.043$, $\lambda^{\prime}=-.122$, $F(1)=.9972$ and $G_{1}(1)=.9992$.

First-order corrections to quantities where they are allowed can be
substantial, but we again find small second-order corrections.  For example,
$F_{1}(1)=1+.345-.005=1.340$ from the zeroth-, first- and second-order
contributions (the unexpanded value is 1.337). The large first-order
corrections in the form factors at zero recoil contain no model dependence
other than the value of
$\overline{\Lambda }$.

The Isgur-Wise function $\zeta$ for baryons is determined in the
model completely by $\alpha=m/\Lambda$.  For $\alpha\approx 1/2$, $\zeta$ is
well approximated for $\omega$ from 1 to 1.5 by  \begin{equation}
\zeta\approx
\left[2/(1+\omega)\right]^{1.32+.70/\omega}. \end{equation}
The slope of this function is $-1.0$ at zero recoil.  We show in Fig. 2
this slope as a function of $\alpha$ and compare it to the slope of the
Isgur-Wise function in the meson case.\footnote{For mesons, $\alpha\equiv
m_{q}/\Lambda$ where $m_{q}$ is the light-quark mass.}  Thus for both
mesons and baryons there is a region of insensitivity to the value of
$\alpha$, and the baryon Isgur-Wise function generally has a less negative
slope.

For the coefficients in the expansion of the squared baryon masses we find
\begin{eqnarray} d &=& .361 + .768\lambda =.394\nonumber \\ d^{\prime}&=&
.038 + .481\lambda + .055\lambda^{2} + .768 \lambda^{\prime}
=-.035.\label{ddp}
\end{eqnarray}  In \cite{FalkNeu}, a first-order mass parameter $\Delta
m_{\Lambda}^{2}$ was defined as \begin{equation}
M=m_{Q}+\overline{\Lambda}+\Delta m_{\Lambda}^{2}/2m_{Q}. \end{equation}  It
turns out to be very small for our choice of $\lambda$,
\begin{equation}\Delta m_{\Lambda}^{2}=(-.043+1.24\lambda=.010)\;{\rm
GeV}^{2}.\end{equation}

We have noted earlier that $\chi (1)=0$.  From (\ref{2ndorder}) this implies
that all first-order corrections to the form factors at zero recoil are
independent of $\lambda$. At second order, the quantities appearing in the
expansions (\ref{2ndorder}) of the form factors have the following
zero-recoil values:
\begin{eqnarray} b_{1}(1)&=&-.008-.534\lambda+5.45\lambda^2 =-.021\nonumber
\\ b_{2}(1)&=&.034+.251\lambda =.045\nonumber \\
b_{3}(1)&=&.040+1.07\lambda-10.9\lambda^2 =.066\nonumber \\
b_{4}(1)&=&-.045-1.49\lambda \nonumber =-.109\\ b_{5}(1)&=&-.034-.745\lambda
=-.066 \nonumber \\ b_{6}(1)&=&-.193-.745\lambda =-.225 \label{a2}.
\end{eqnarray} The pattern observed in the first-order corrections
continues; the second-order corrections depend on $\lambda$ but now there is
no dependence on the second-order parameter $\lambda^{\prime}$.  We display
$\chi$ and $b_{i}$ as functions of $\omega$ in Fig. 3.

There are interesting parallels between these results and the meson case, in
which quantities corresponding to $\lambda$ were labeled $g$ and $h$.  The
parameter $h\approx.25$, which was fixed by hyperfine mass splitting, is
significantly larger than $\lambda$.  We argued that $h^2$ terms, analogous
to the
$\lambda^2$ terms in $b_{1}(1)$ and
$b_{3}(1)$, must produce substantial corrections to ${h}_{+}(1)$ and
${h}_{{A}_{1}}(1)$, which are quantities analogous to $F(1)$ and
$G_{1}(1)$.  The large coefficients of
$\lambda^2$ appearing explicitly in (\ref{a2}) support this interpretation.

Neglecting the lepton mass, the differential spectrum for the unpolarized
$\Lambda_{b}\rightarrow
\Lambda_{c}\ell^{-}\overline{\nu}$ decay is given by
\begin{eqnarray}
d\Gamma/d\omega&=&\kappa\sqrt{\omega^2-1}\left[(\omega-1)
\left\{ 2\widehat{q}^{2}F_{1}^{2} + [(1+r)F_1+(\omega+1)(rF_2+F_3)]^2
\right\} \right. \nonumber \\ & & +\left. (\omega+1)
\left\{ 2\widehat{q}^{2}G_{1}^{2} + [(1-r)G_1-(\omega-1)(rG_2+G_3)]^2
\right\} \right],
\end{eqnarray} where $\kappa=G_{F}^{2}|V_{cb}|^{2}M^{5}r^{3}/24\pi^{3}$,
$r=M^{\prime}/M$ and $\widehat{q}^{2}\equiv q^{2}/M^{2}=1+r^2-2r\omega$.  It is
given in the heavy-quark limit by
\begin{equation}
d\Gamma/d\omega=\kappa\sqrt{\omega^{2}-1}\left[
6\omega\widehat{q}^2+4r(\omega^2-1)\right]\zeta(\omega)^2.
\end{equation}

For the decay rate we find \begin{equation}
\Gamma=.65\times10^{11}(|V_{cb}|/.04)^{2} s^{-1} \end{equation} from the
unexpanded model. The rate is reduced by $2$\%  when the form factors are
replaced by their values in the heavy-quark limit, the large first-order
corrections to the form factors having largely cancelled.  The rate is quite
stable in the scalar mass; we find only a 10\% increase even when the
scalar mass is allowed to vanish.  The rate changes by about 10\% when the
$b$ and $c$ quark masses are varied by $\pm 50$ MeV.  However, our
analysis of mesons has indicated that the allowed variation of these
masses within the context of our model is much smaller, on the order of $\pm
10$ MeV, and this reduces the variation in the rate to a few percent.  The
main point is that the process $\Lambda_{b}\rightarrow
\Lambda_{c}\ell^{-}\overline{\nu}$ is free of the potentially large
$1/{m}_{Q}^{2}$ corrections which plague heavy-meson decay, and we therefore
encourage efforts to measure this process as a precise determination of
$|V_{cb}|$.

\vspace{2ex}

\noindent {\bf ACKNOWLEDGMENT}
\vspace{1ex}

We thank M. Luke for discussions.  This research was supported in part by the
Natural Sciences and Engineering Research Council of Canada.

\newpage \noindent{\bf FIGURE CAPTIONS} \vspace{6ex}

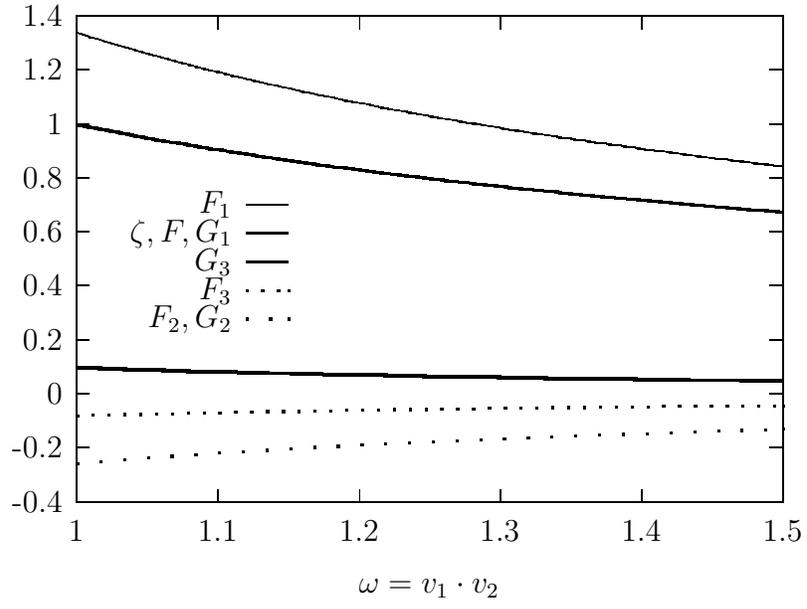
\noindent {\bf FIG. 1:} Form factors in the unexpanded model with
$m_{b}=4.85$ GeV, $m_{c}=1.50$ GeV and $m=500$ MeV.
$F\equiv F_{1}+F_{2}+F_{3}$ and
$G_{1}$ are indistinguishable from the model's Isgur-Wise function
$\xi$ on the scale of this plot.
\vspace{4ex}

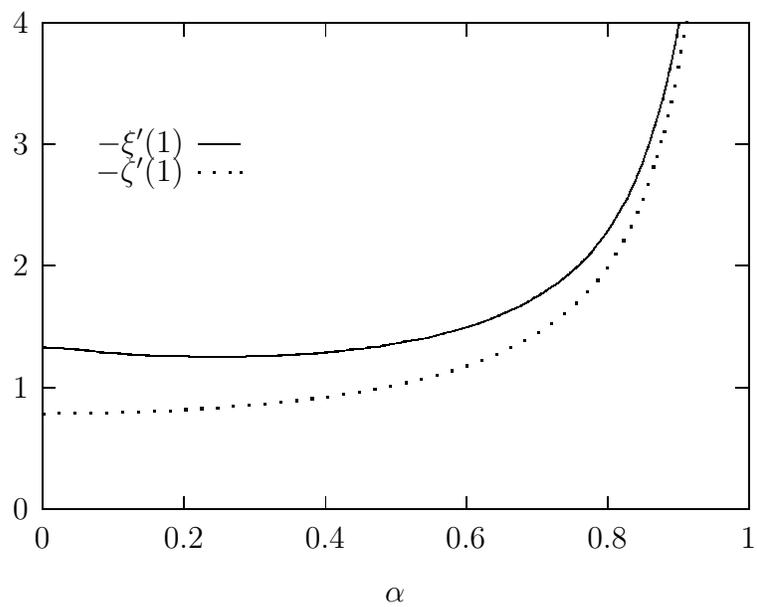
\noindent {\bf FIG. 2:} Comparison of zero-recoil slopes of Isgur-Wise
functions $\xi$ for mesons and $\zeta$ for baryons.  For mesons,
$\alpha\approx .37$ while for baryons $\alpha\approx .49$.
\vspace{4ex}

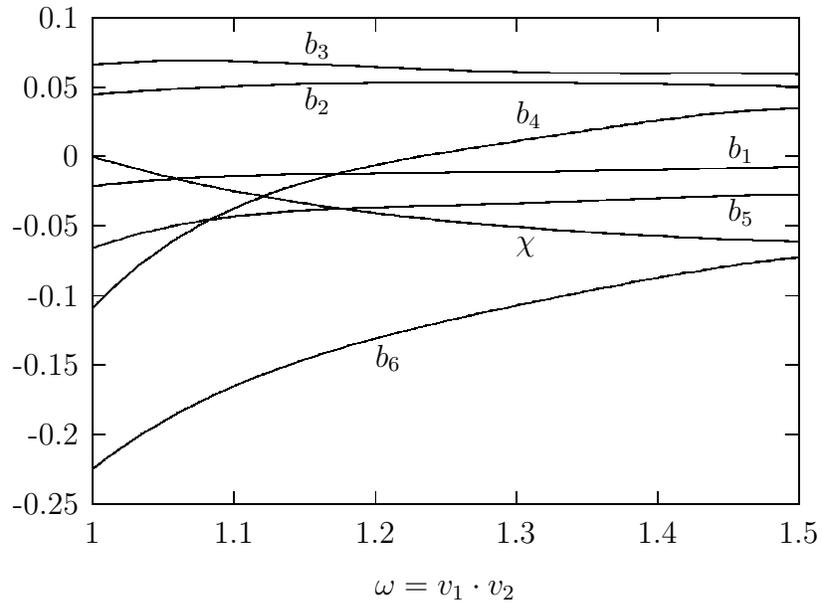
\noindent {\bf FIG. 3:} First-order universal function $\chi$ and
second-order functions $b_{i}$ for $\Lambda=1.019$ GeV, $\lambda=.043$ and
$\lambda^{\prime}=-.122$.

\end{document}